# Towards a Taxonomy of Large Language Model based Business Model Transformations

Jochen Wulf[1] and Jürg Meierhofer[1]

[1] Zurich University of Applied Sciences,
Technikumstrasse 81, 8401 Winterthur, Switzerland

**Abstract.** Research on the role of Large Language Models (LLMs) in business models and services is limited. Previous studies have utilized econometric models, technical showcases, and literature reviews. However, this research is pioneering in its empirical examination of the influence of LLMs at the firm level. The study introduces a detailed taxonomy that can guide further research on the criteria for successful LLM-based business model implementation and deepen understanding of LLM-driven business transformations. Existing knowledge on this subject is sparse and general. This research offers a more detailed business model design framework based on LLM-driven transformations. This taxonomy is not only beneficial for academic research but also has practical implications. It can act as a strategic tool for businesses, offering insights and best practices. Businesses can leverage this taxonomy to make informed decisions about LLM initiatives, ensuring that technology investments align with strategic goals.

## 1 Introduction

Large Language Models (LLMs) are extremely large-scale artificial neural networks that are trained with terabytes of textual content to complete texts. LLMs can be used for a variety of purposes, such as text summaries, sentiment analysis, or named entity recognition. The expected economic potential of generative AI applications, and LLMs in particular, is exceptional. A study by McKinsey & Company (Chui et al. 2023) estimates that generative AI could add between $2.6 to $4.4 trillion annually to the global economy, potentially doubling if integrated into existing software. Around 75% of this value according to this study is concentrated in customer operations, marketing and sales, software engineering, and R&D, with applications like customer interaction support and content generation. Industries like banking, high tech, and life sciences are predicted to see significant impacts, with potential values reaching hundreds of billions annually. Most notably, generative AI can automate a substantial portion of tasks that currently occupy employees, especially in knowledge work that requires understanding natural language (Davenport and Mittal 2022). In an extreme scenario, the adoption of generative AI could lead to half of today's work activities being automated by 2045, boosting labor productivity, but necessitating investments in worker training and transition support (Chui et al. 2023).

Already today, a vast and very dynamic ecosystem of LLM software and services is emerging. Vendors across the enterprise stack are swiftly adapting to market changes, with both established platforms integrating GenAI and new generative platforms emerging (Miclaus, Radu 2023). But LLMs also have an impact on incumbent firms. Many service industries are predicted to undergo significant transformations driven by LLM innovations. In the manufacturing industry, for example, LLMs can contribute to guided machine maintenance, materials processing efficiency, and industrial equipment longevity (Eichhorn, Ellen 2023).

According to a Gartner analysis, LLM productivity assistants like Microsoft 365 Copilot and Google Workspace offer quick benefits with easy adoption but may lose competitive edge over time (Sallam, Plath, and Zijadic, 2023). Integrating these tools with other business processes can offer a sustained competitive advantage, especially when embedded in domain-specific applications, though at higher costs and risks. Achieving a unique competitive edge with LLMs demands investment in transformational use cases, accepting higher costs, complexities, and risks, prioritizing strategic value over specific productivity gains (Sallam, Plath, and Zijadic, 2023).

Although there are already initial indications of the productivity gains that can be achieved through the use of LLMs, the role of this technology for business model innovation is still largely unclear from an academic perspective (Brynjolfsson, Li, and Raymond 2023; Kanbach et al. 2023). In this article, we address this research gap and discuss the results of a firm-level study of LLM adoption. We present a taxonomy of mechanisms of how LLMs transform business models. This typology results from a structured analysis of 50 real-world use cases.

## 2 Theoretical Background

### 2.1 Business Model Transformation

In this work, we use Osterwalder´s business model canvas to structure the transformative impact of LLMs on business models (Osterwalder 2004). A business model is a theoretical framework that outlines the various components and how they interact, detailing how a company makes profits. It describes the value a business provides to specific customer groups, as well as the structure of the company and its collaboration with partners to produce, promote, and deliver this value. The goal is to achieve consistent and profitable revenue (Massa, Tucci, and Afuah 2017).

The business model canvas is a strategic management tool used to develop and document business models for product and service firms alike by visualizing key elements such as the offering, infrastructure, customers, and finances. It consists of the following building blocks (Osterwalder 2004):

- Value Proposition: A Value Proposition provides a comprehensive perspective on the array of products and services a company offers that benefit the customer.
- Target Customer: The Target Customer refers to the specific group of consumers a company aims to serve and provide value for.
- Distribution Channel: A Distribution Channel is a method used to connect with and reach the customer.
- Relationship: The Relationship outlines the type of connection a company forms with its customers.
- Value Configuration: The Value Configuration outlines how activities and resources are organized to deliver value to the customer.
- Capability: A capability refers to the consistent set of actions a company can perform to deliver value to its customers.
- Partnership: A Partnership is a collaborative arrangement willingly formed between multiple companies with the aim of delivering value to the customer.
- Cost Structure: The Cost Structure depicts the monetary breakdown of all the resources utilized in the business model.
- Revenue Model: The Revenue Model outlines how a company generates income from various revenue streams.

A business model transformation can be defined as "a change in the perceived logic of how value is created by the corporation, when it comes to the value-creating links among the corporation's portfolio of businesses, from one point of time to another." (Aspara et al. 2013). In this article, we focus on business model transformation, rather than business model innovation, because we want to analyze the transformative impact of LLMs on incumbent firms. While LLMs also enable completely new services for startups and new ventures, this will not be discussed in this article.

### 2.2 The Economic Value of Large Language Models

LLMs have emerged as powerful artificial intelligence algorithms for language understanding and generation (Zhao et al. 2023). LLMs are developed through pre-training transformer models over large-scale corpora, leading to significant advancements in natural language processing (NLP) tasks (Vaswani et al. 2017). Scaling LLMs to larger parameter sizes has shown improved performance and unique abilities. This particularly applies to the ability to learn in-context and achieve enhanced performance, demonstrating superior results in a wide range of natural language processing tasks, including language understanding, generation, information retrieval, and other NLP applications (Liu et al. 2023).

Prior literature on the economic value of LLMs in companies is very scarce. Some articles discuss the impact of LLMs on the individual worker level. Brynjolfsson et al (2023), for example, study the impact of LLM on the work of 5,179 customer support agents. The introduction of a conversational assistant increased productivity by 14%, benefiting novice and low-skilled workers the most. The AI tool also improved customer sentiment, reduced managerial intervention requests, and enhanced employee retention. Noy and Zhang (2023) conduct an online experiment among college-educated professionals and show a substantial productivity increase in general writing tasks. In a study by Peng et al. (2023), 95 programmers test the efficiency of Copilot in creating a JavaSript HTTP server. Those using Copilot finish 55.8% faster than the control group, with a subsequent increased willingness to pay for the tool. Chen et al. (2023) use an exposure scoring method to model exposure to LLM capabilities. Their findings suggest that higher-paying and experience-intensive jobs face greater displacement risks.

Regarding the impact of LLMs on the firm level, some researchers use self-developed models to demonstrate the LLM potential. Chen et al. (2023), for example, showcase the ability of LLMs to support decision making in financial markets by processing text from corporate financial statements and correlating corporate sentiments to stock return performance. Musser (2023) introduces an econometric model and demonstrates via simulations that LLMs may result in up to 70% reduction of content generation costs for influence operations.

One team of authors has conducted a literature analysis to analyze how generative AI impacts business model innovation (Kanbach et al. 2023). While this study does not focus on LLMs but also discusses text-to-image applications, it provides first valuable insights. Most notably, the authors distinguish between three categories of business model innovation: value creation innovation, new proposition innovation, and value capture innovation. This research question, however, requires further investigations because empirically grounded and finer grained insights into business model impacts are still missing.

In summary, there is an apparent lack of empirical work on firm-level use and effects of LLMs. The question of how LLMs influence business models has not been approached empirically.

## 3  Research Methodology

A taxonomy is a "system of grouping objects of interests in a domain based on common characteristics" that are derived conceptually or empirically (Nickerson, Varshney, and Muntermann 2013; Bailey 1994). As discussed in prior management literature, it can help identify interrelationships between design dimensions and thus support the design decisions in domains such as business models and organizational processes (Van Giffen, Herhausen, and Fahse 2022; Möller et al. 2022; Staub et al. 2021; Weking et al. 2020).

We used the method for taxonomy development defined by Nickerson et al (Nickerson, Varshney, and Muntermann 2013), who describe an iterative methodology consisting of empirical-to-conceptual and conceptual-to-empirical iterations of data analysis. In the empirical-to-conceptual iterations, the researcher inductively identifies common characteristics of sample objects. In the conceptual-to-empirical iterations, the researcher deductively defines taxonomy's dimensions and then classifies real-world objects. We classify the use cases based on common characteristics derived from the meta-characteristic (particularly the business model dimensions), forming initial dimensions for the taxonomy. The process iterates, offering new perspectives while predetermined ending conditions are assessed. The individual steps of the taxonomy development process are summarized in Table 1. We used three empirical-to-conceptual and one conceptual-to-empirical iterations with 15-20 additional use cases each.

**Table 1.** Taxonomy Development Summary

| Step | Result |
|---|---|
| 1. Meta-characteristic | Characteristic business model features |
| 2. Empirical to conceptual | Desirability components: value proposition, distribution channel |
| 3. Empirical to conceptual | Feasibility components: value configuration, capability |
| 4. Conceptual to empirical | Viability components: revenue model, cost structure |
| 5. Empirical to conceptual | No new dimensions |
| 6. Ending conditions | No new dimensions in the last iteration, all sample objects examined, concise, robust |

*Step 1: Meta-characteristics*: Our taxonomy aims to systematically categorize different examples and interpretations of LLM-based business transformations from a business model viewpoint. As such, we identify the primary business model configurations as the foundational characteristic that will guide the selection of design dimensions throughout our taxonomy creation process.
*Step 2: Empirical to conceptual iteration*: Because domain knowledge is limited, we start with an empirical-to-conceptual iteration. We group a set of randomly selected use cases with respect to common desirability characteristics.
*Step3: Empirical to conceptual iteration*: Next, we inductively identify feasibility characteristics by adding further use cases and grouping them with regard to value configuration and capability aspects.
*Step4: Conceptual to empirical iteration*: In order to get a different perspective on the taxonomy, we deductively define viability characteristics and classify the collected set of use cases.
*Step3: Empirical to conceptual iteration*: We collect a last set of use cases and apply the taxonomy classification scheme. Since no new dimensions emerge, we check the ending conditions.
*Step 6: Ending conditions:* We used both subjective and objective criteria to determine the acceptance of our taxonomy. On an objective level, the taxonomy consists of dimensions that don't overlap and cover all necessary areas. It encompasses the primary facets of a business model transformation, such as reconfigured value proposition, channels, value configurations or capabilities with resulting effects on the cost or revenue models, as defined in the prior chapter. From a subjective standpoint, the taxonomy is considered clear, adaptable, insightful, and all-encompassing.

Our data collection approach was as follows. We used the search term "'Large Language Model' OR LLM OR GPT" AND "'Case Study' OR 'Use Case' OR 'Case Example' OR Application" in order to identify use cases via the search engines Google, Bing and scholar.google.com. We excluded use cases without clear evidence of LLM implementation. The data collection resulted in 50 use cases. A list of the use cases and associated data sources can be provided upon request. We show the distribution of industry sectors in Figure 1.

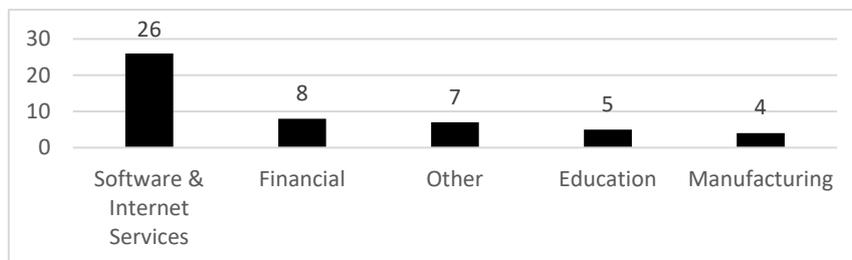

**Fig. 1.** Number of use cases per sector

The majority of use case companies are from the Software & Internet Services sector. A notable number of companies further belong to the Financial, Education, and Manufacturing sectors. The sector class Others includes Health, Defense and Airline industries among others. The distribution of company sizes is shown in Figure 2.

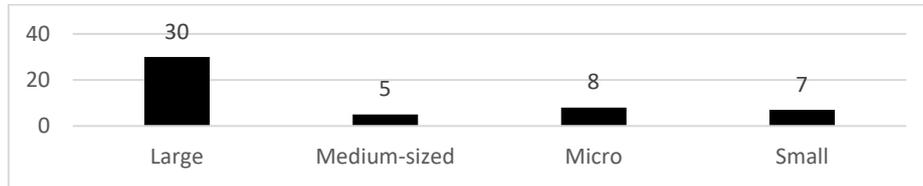

**Fig. 2.** Company sizes of the use case companies.

The majority are large firms with over 250 employees and over €43 million turnover. However, there are also micro firms (<10 employees), small firms (<50 employees), and medium-sized firms (<=250 employees) in the data set.

For inductive and deductive classification, we used a coding scheme that identifies characteristic business model features in the use case descriptions. Table 2 lists coding examples for the different business model components.

**Table 2.** Coding Examples.

| Business Model Component | Exemplary Text Extract | Number of Use Cases |
|---|---|---|
| Value Proposition | "**dietary considerations**: … users can ask questions like "What kind of side dishes should I serve with lamb chops?" | 20 |
| Distribution Channel | "use ChatGPT to **improve the customer experience on its website**…to improve the FAQ section" | 6 |
| Value Configuration | "automated system **manages requests and cancellations**, and **creates draft responses** to guests' reviews, addressing feedback with a full understanding of specific positive and negative elements" | 10 |
| Capability | "**finding and simplifying information** buried within hundreds of thousands of complex documents." | 14 |
| Cost model | "Rolls Royce is using AI and Natural Language Processing (NLP) to **reduce costs**" "**efficient** way to learn about judicial processes" | 24 |
| Revenue model | "**Pricing: $19** per user per month" "Pro plan gives you unlimited projects, words, and brand voices: **$49/mo**" | 26 |

## 4 Results

Four archetypes of how LLMs impact business model transformation emerge from our data: 1) new customer benefits, 2) new sales and communication channels, 3) increased business process automation and 4) improved use of information resources. While, from a business model viability perspective, archetypes 1 and 2 relate to revenue increases, archetypes 3 and 4 address efficiency increases. In the following, we discuss the four archetypes as well as their performance impact.

### 4.1 New customer benefits

We identified four mechanisms of how LLMs contribute to the development of novel customer-facing value propositions: Personal Assistance, Coaching, Content Generation, and Speech Interaction (see Table 3).

**Table 3.** Value Proposition Dimension.

| Personal Assistance | Coaching | Content Generation | Speech Interaction | Number of Use Cases |
|---|---|---|---|---|
| X | - | - | - | 6 |
| - | X | - | - | 4 |
| - | - | X | - | 7 |
| - | - | - | X | 3 |

*Personal Assistance* refers to scenarios in which LLMs help to provide personalized information to users during service consumption. The grocery delivery platform Instacart, for example, uses LLMs to provide personalized grocery recommendations by answering questions such as "What are good sauces for grilling chicken?".

*Coaching* addresses use cases, in which LLMs supports individual learning paths in education. Khan Academy, for example, automatically detects errors in programming tasks and assists users in finding these errors without directly naming them.

*Content Generation* encompasses use cases, where textual components of a value propositions are created with LLMs. The car sales platform carmax, for example, uses LLMs to produce textual descriptions of the different car models and a two-sentences summary of customer reviews.

*Speech Interaction* covers use cases in which LLMs support voice-based machine interactions. The car manufacturer Mercedes, for example, integrates LLMs into the car infotainment system. This allows to have conversations and answer complex driver questions, for example about a driver's destination.

### 4.2 New sales and communication channels

We further discovered two mechanisms of how LLMs enable novel distribution channels: Presales Automation and Customer Service Automation (see Table 4).

**Table 4.** Distribution Channel Dimension.

| Presales Automation | Customer Service Automation | Number of Use Cases |
|---|---|---|
| X | - | 2 |
| - | X | 4 |

*Presales Automation* addresses scenarios in which LLMs help to provide personalized information to prospective customers. Air India, for example, employs an LLM-powered chatbot to answer questions on information that is publicly available on their website. This includes questions about booking, flight cancellations, the baggage policy and the loyalty program.

*Customer Service Automation* deals with handling customer inquiries in the after sales phase. The non-profit foundation Solana Foundation, which supports open cryptocurrency, uses LLMs for customer onboarding. The LLM-powered chatbot allows users to list NFTs, move tokens, view transactions, analyze data, and search for NFT collections based on their floor price.

### 4.3 Increased business process automation

Regarding the Value Configuration component of business models, our data revealed three mechanisms of how LLMs contribute to an increased automation of information-intensive business processes: Front Office Process Automation, Back Office Process Automation, and Software Development Automation (see Table 5).

Table 5. Value Configuration Dimension.

| Front Office Process Automation | Back Office Process Automation | Software Development Automation | Number of Use Cases |
|---|---|---|---|
| X | - | - | 6 |
| - | X | - | 3 |
| - | - | X | 1 |

*Front Office Process Automation* encompasses scenarios in which LLMs support the automation of customer communication. Radisson Hotel Group, for instance, uses an LLM-supported system that handles guest cancellations and crafts preliminary replies to guest reviews, comprehensively addressing feedback by pinpointing both positive and negative aspects.

*Back Office Process Automation* deals with automating company-internal tasks related to handling a customer order. Swiss retail bank Migros Bank, for example, employs LLMs to automate mortgage application processes. The system verifies the applicant's income and discerns the necessary documents based on the marital status. This goes beyond the standard RPA and OCR, delving deep into the nuances and specific needs of each individual case.

*Software Development Automation* addresses an increased efficiency in code development processes through code completion, code documentation or automated debugging. At the language app provider duolingo, for example, the use of LLMs enhances software developer efficiency by minimizing context shifts, lessening the demand for hand-written basic code, and thereby allowing developers to concentrate on addressing intricate business issues.

### 4.4 Improved use of information resource

Regarding the Capability business model component, we discover that LLMs can strongly improve the utilization of information resources. This includes a facilitation of information access (Information and Knowledge Management) as well as Information Extraction (see Table 6).

Table 6. Capability Dimension.

| Information and Knowledge Management | Information Extraction | Number of Use Cases |
|---|---|---|
| X | - | 7 |
| - | X | 7 |

*Information and Knowledge Management* refers to scenarios in which LLMs enables quick access to information from vast amounts of textual data. The investment firm Morgan Stanley, for example, offers an LLM-powered chatbot to its employees. This chatbot is designed to thoroughly search wealth management content, essentially tapping into the collective wisdom of Morgan Stanley Wealth Management.

*Information Extraction* goes a step further by aggregating information that is embedded in textual documents of deducing novel insights from this data. Zurich Insurance, for example, uses LLMs to analyze whether client claims are covered by an insurance contract. LLMs allow them to scrutinize contract terms to identify potential areas of differing interpretations between them and their clients.

### 4.5 Business Model Viability

When looking at the revenue and cost models, our data reveals two mechanisms of how LLMs impact business model viability: Efficiency Increase and Revenue Increase (see Table 7).

**Table 7.** Key Resource Dimensions.

| Efficiency Increase | Revenue Increase | Number of Use Cases |
|---|---|---|
| X | - | 24 |
| - | X | 26 |

*Efficiency Increase* refers to the reduction of manual work owing to the (semi-)automation of textual data or code processing. This performance effect was relevant for all use cases relating to business process automation and the improved use of information resources. At duolingo, the use of LLMs for software development led to an estimated increase of 25% in developer speed, and a 10% increase for those already familiar with that same codebase. Github, the provider of the LLM-powered pair programming service Github Copilot, found in a survey among its customers that LLM users complete software development tasks 55% percent faster than developers without Copilot support (Kalliamvakou 2022).
*Revenue Increase* relates to additional profits with new offerings or due to customer growth. This performance effect was relevant for the use cases in the areas of new customer benefits and new sales and communication channels. One example is Khan Academy with its LLM-powered tutor Khanmigo. This offering comes with a dedicated pricing model of $9 per month or $99 per year.

## 5 Discussion

### 5.1 Contribution to Theory

Socio-technical research on how LLMs contribute to business models and services is still scarce. Prior works, as discussed in section 2.2, use econometric models and simulations (Musser 2023), a technical showcase (B. Chen, Wu, and Zhao 2023) or literature analysis (Kanbach et al. 2023). This research, to the best of our knowledge, is the first empirical approach that studies the business model impact of LLMs on the firm level.

In this research we present a taxonomy that informs further research which may further elaborate on the success criteria for implementing LLM-based business models and deepen the knowledge on LLM-based business model transformations. Prior knowledge on this topic is anecdotical and remains on a very coarse level (Kanbach et al. 2023). With the presented archetypes for LLM-driven business transformations, we introduce a finer-grained business model design conceptualization. Our results may, for example, inform further research on LLM-driven business models in particular industries.

### 5.2 Contribution to Practice

The developed taxonomy also contributes to practice in that it may serve as a strategic tool. Business model taxonomies provide practitioners with inspirations and best practice examples – practitioners can use taxonomies to compare and innovate their own business (Möller et al. 2022).

Companies can use our taxonomy for an informed decision making on LLM initiatives and to align technology investments with strategic objectives. The taxonomy will help practitioners to develop their dynamic capabilities and become more agile regarding AI-related innovations.

## 6 Conclusion

In this research, we present an empirical approach towards developing a taxonomy on LLM-based business model transformation. We introduce four archetypes and discuss how to configure the business model components value proposition, customer channel, value configuration, capability, cost model, and revenue model.

As with all research endeavors, some limitations need to be considered. First, our results may be impacted by our data collection strategy. Even though we adopted a wide approach towards use case identification and used several search engines, the inclusion and emergence of additional LLM use cases may create the need to further develop the introduced taxonomy. Second, owing to the novelty of LLMs, viability categorization was limited to expressed goals, such as efficiency gains or revenue increase. Further research is needed to quantify the economic value generated via the business model transformation archetypes. Third, this study is limited to the transformation of incumbents' business models. The question of how LLMs enable the design and implementation of green-field business models remains to be explored by further research.